\documentclass[numreferences]{crckbked}

\makeindex
\usepackage{times}
\usepackage{bm}
\usepackage{klucite}
\usepackage{graphicx} 

\newcommand{\pdag}{{\phantom{\dagger}}}

\begin{document}

\begin{opening}
\title{Coulomb blockade and Kondo effect in quantum dots}

\author{L.I. \surname{Glazman}$^1$}
\author{M. \surname{Pustilnik}$^2$}

\institute{
$^1$Theoretical Physics Institute, University of Minnesota, 
\\ Minneapolis, MN 55455 \\
$^2$School of Physics, Georgia Institute of Technology, 
\\ Atlanta, GA 30332 
}

\runningtitle{Coulomb blockade and Kondo effect}
\runningauthor{L.I. Glazman and M. Pustilnik }

\begin{abstract}
We review the mechanisms of low-temperature electron transport across 
a quantum dot weakly coupled to two conducting leads. Conduction in 
this case is controlled by the interaction between electrons. At temperatures 
moderately lower than the single-electron charging energy of the dot, the 
linear conductance is suppressed by the Coulomb blockade. Upon further 
lowering of the temperature, however, the conductance may start to increase 
again due to the Kondo effect. This increase occurs only if the dot has a 
non-zero spin $S$. We concentrate on the simplest case of $S=1/2$, and 
discuss the conductance across the dot in a broad temperature range, 
which includes the Kondo temperature. Temperature dependence of the 
linear conductance in the Kondo regime is discussed in detail. We also 
consider a simple (but realistic) limit in which the differential conductance 
at a finite bias can be fully investigated. 
\end{abstract}

\end{opening}

\section{Introduction}

In this review, we define a quantum dot as a small paddle of electron 
liquid which is attached by tunnel junctions to two massive conductors 
(leads). The number of electrons on the dot in equilibrium $N$ can be 
controlled by varying the voltage applied to a gate -- an auxiliary 
electrode coupled to the dot only capacitively (no tunneling occurs 
between the dot and gate). If the conductances of the lead-dot junctions 
are small compared to the conductance quantum $e^2/h$, and if the 
temperature $T$ is low enough, then $N$ is an integer at almost any gate 
voltage $V_g$. Exceptions are small intervals around the discrete set of 
values of $V_g$, at which an addition of a single electron to the dot 
does not change the electrostatic energy. Such a degeneracy between 
different charge states of a quantum dot allows for an activationless 
electron transfer through it, whereas for all other values of $V_g$ the 
activation energy for the conductance $G$ across the dot is finite. 
The resulting oscillatory dependence $G(V_g)$ is the hallmark of the 
Coulomb blockade phenomenon~\cite{blockade}. The contrast between the 
low- and high-conductance regions (Coulomb blockade valleys and peaks, 
respectively) gets sharper at lower temperatures. The described
pattern of the $G(V_g,T)$ dependence is clearly observed down to the
lowest attainable temperatures in experiments on tunneling through
small metallic islands~\cite{Devoret97}
However, small quantum dots formed in GaAs heterostructures display 
a different behavior~\cite{kondo_exp}: in some Coulomb blockade valleys 
the dependence $G(T)$ is not monotonic and has a minimum at a finite 
temperature. This minimum is similar in origin~\cite{kondo_popular} 
to the well-known non-monotonic temperature dependence of the resistivity 
of a metal containing magnetic impurities~\cite{Kondo} -- the Kondo 
effect.

We present here an introduction to the theory of Kondo effect in
quantum dots, concentrating on the so-called {\it Constant Interaction
Model}. Despite its simplicity, the model allows one to describe the
main conduction mechanisms, including the most important version of
the Kondo effect in quantum dots.

\section{Constant Interaction Model} 
\label{CIM}

The Hamiltonian of interacting electrons confined to a quantum dot has 
the following general form:
\begin{equation}
{\hat H}_{dot} = 
\sum_{ij;s}
{\cal H}^\pdag_{ij}d^\dagger_{is}d^\pdag_{js}
+
\frac{1}{2}\sum_{s s'}\sum_{ijkl} {\cal H}^\pdag_{ijkl}
d^\dagger_{i s} d^\dagger_{j s'} 
d^\pdag_{k s'}d^\pdag_{l s}.
\label{2.1}
\end{equation}
Here an operator $d^\dagger_{is}$ creates an electron with spin $s$ in
the orbital state $\phi_i({\bf r})$;
${\cal H}_{ij}={\cal H}^*_{ji}$ is an Hermitian matrix describing
the single-particle spectrum, and the matrix elements ${\cal
H}_{ijkl}$ depend on the electron-electron
interaction potential $V({\bf r}_1-{\bf r}_2)$ and on the chosen basis
of orbital states:
\begin{equation}
{\cal H}_{ijkl}
=\int d{\bf r}_1 d{\bf r}_2 V({\bf r}_1-{\bf r}_2)
\phi_{i}({\bf r}_1)
\phi_{j}({\bf r}_2)
\phi_{k}^*({\bf r}_2)
\phi_{l}^*({\bf r}_1).
\label{2.2}
\end{equation}
Further simplification of the Hamiltonian (\ref{2.1}) is possible if the 
following conditions are met: \\
({\it i}) the electron-electron interaction within the dot is not too 
strong (the gas parameter characterizing the electron-electron 
interaction in a non-ideal Fermi gas $r_s\lesssim 1$) so that the Fermi 
liquid description is applicable; \\
({\it ii}) the quasiparticle spectrum is not degenerate
near the Fermi energy (which is satisfied, in general, if the motion of 
electrons within the dot is chaotic); \\
({\it iii}) the dot is in the metallic regime, {\sl i.e.} the Thouless 
energy of the dot $E_T$ exceeds the mean single-particle level
spacing $\delta E$.

The Thouless energy $E_T$ is well-defined if the electron motion within 
it is chaotic due to either disorder or irregular shape of the dot. For 
a dot of linear size $L$ it equals
\begin{equation}
E_T \simeq \left\{
\begin{array}{lr}
\hbar D/L^2, & ~l\ll L \\
{\hbar v_F}/{L}, & ~l\gg L
\end{array}
\right.,
\label{2.3}
\end{equation}
where $l$ if the elastic mean free path and $D$ is the diffusion constant
in a disordered dot. The average level spacing $\delta E$ can be estimated 
as
\begin{equation}
\delta E \simeq 1/\nu_dL^d.
\label{2.4}
\end{equation}
Here $d$ is the dimensionality of the system ($d=2$ for a quantum dot 
formed in a two-dimensional electron gas in a semiconductor 
heterostructure, and $d=3$ for a metallic nanoparticle); $\nu_d$ is the 
density of states of the material. For a ballistic electron motion 
($l\gg L$) the ratio 
\[
E_T/\delta E \sim L/\lambda_F\sim N^{1/d}
\] 
is large for large number of electrons $N$ on the dot.

If the conditions ({\it i})-({\it iii}) are met, then the Random
Matrix Theory (RMT) is a good starting point for describing
non-interacting quasiparticles, see~\cite{Beenakker} for a review; 
the matrix elements ${\cal H}_{ij}$ belong to a Gaussian 
ensemble~\cite{Mehta}. The matrix elements do not depend on spin, 
and therefore each eigenvalue $\epsilon_n$ of the matrix ${\cal H}_{ij}$ 
represents a spin-degenerate energy level. The spacings 
$|\epsilon_{n+1}-\epsilon_n|$ between consecutive levels obey the 
Wigner-Dyson distribution~\cite{Mehta}; the average value of 
$|\epsilon_{n+1}-\epsilon_n|$ is $\delta E$.

It turns out~\cite{RMT} that the vast majority of the matrix elements 
${\cal H}_{ijkl}$ are small. In the limit $E_T/\delta E\to\infty$, only 
relatively few ``most diagonal'' elements remain finite. In this limit the 
interaction part $H_{int}$ of the Hamiltonian of a non-superconducting 
dot may be cast in the form~\cite{ABG}:
\begin{equation}
\hat{H}_{int}=E_C\left(\hat{N}-{\cal N}\right)^2 - E_S \hat{\bf S}^2. 
\label{2.5}
\end{equation}
Here 
\begin{equation}
 \hat{N} =\sum_{ns} d^\dagger_{ns}d^\pdag_{ns},
\quad
\hat{\bf S} = \sum_{nss'}
d^\dagger_{ns}\frac{{\bm\sigma}_{ss'}}{2}d^\pdag_{ns'}
\label{2.6}
\end{equation}
are the operators of total number of electrons and total spin on the
dot.  Obviously, the Hamiltonian~(\ref{2.5}) is invariant with respect 
to a change of the single-particle basis $\phi_i({\bf r})$.

\begin{figure}[h]
\centerline{\includegraphics[width=0.45\textwidth]{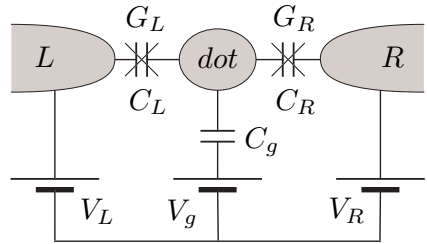}}
\vspace{2.5 mm}
\caption{Equivalent circuit of a dot connected to two leads by tunnel
  junctions (with conductances $G_L$ and $G_R$) and coupled
  via a capacitor (with capacitance $C_g$) to the gate. The total
  capacitance of the dot $C=C_L+C_R+C_g$.}
\label{circuit}
\end{figure}

The first term in Eq.~(\ref{2.5}) represents the electrostatic energy. 
In the conventional equivalent circuit picture, see Fig.~\ref{circuit}, 
the charging energy $E_C$ is related to the total capacitance $C$ 
of the dot, $E_C=e^2/2C$, and the dimensionless parameter ${\cal N}$ 
is proportional to the gate voltage, ${\cal N}=C_gV_g/e$, where $C_g$ 
is the capacitance between the dot and the gate.  For a mesoscopic 
($\lambda_F\ll L$) conductor, the charging energy is large compared 
to the level spacing $\delta E$. Indeed, using the estimate $C\sim\kappa L$, 
where $\kappa$ is the dielectric constant, we find
\begin{equation}
\frac{E_C}{\delta E} 
\sim
\frac{e^2}{\kappa\hbar v_F} \left(\frac{L}{\lambda_F}\right)^{d-1}
\sim~
r_s N^{d-1},
\label{2.7A}
\end{equation}
where $v_F$ is the Fermi velocity, and $r_s$ is the conventional gas 
parameter. Except an exotic case of an extremely weak interaction, having 
a large number of electrons $N \sim (L/\lambda_F)^d\gg 1$ on a quantum dot 
guarantees that the inequality $E_C\gg \delta E$ is satisfied in dimensions 
$d=2,3$. For the smallest quantum dots formed in GaAs heterostructures, 
$E_C/\delta E\sim 10$~\cite{kondo_exp}; for a metallic nanoparticle with 
$L\simeq 5$~nm this ratio is about $100$, see, {\sl e.g.},~\cite{Davidovic}. 
We should mention also that in the $d=1$ case, the estimate (\ref{2.7A}) is 
of little help, as even a weak electron-electron interaction results in a 
formation of a Luttinger liquid~\cite{MPAFrev}, significantly affecting 
the nomenclature of the elementary excitations. In the case of a single-mode 
finite-length Luttinger liquid, the elementary excitations can be viewed 
as $1d$ plasmon waves in a confined geometry, so the corresponding level 
spacing is $\delta E\sim E_C$.

The second term in Eq.~(\ref{2.5}) is the exchange interaction
characterized by the exchange constant $E_S$.  This constant is small 
in the case of weak electron-electron interaction: $E_S\sim r_s\delta E$ 
(there is an additional $\ln (1/r_s)$ factor in this estimate in the case 
of a $2d$ dot). Smallness of the ratio $E_S/\delta E$ guarantees the 
absence of a macroscopic (proportional to the volume of the dot) spin in 
the ground state, in accordance with the well-known Stoner criterion for 
the itinerant magnetism~\cite{Ziman}. If all the orbital levels were 
equidistant, then the spin of an even-electron state would be zero, while 
an odd-electron state would have spin $1/2$. However, the level spacings
$|\epsilon_{n+1}-\epsilon_n|$ are random. If the spacing between some 
orbital levels is accidentally small, the dot may acquire a spin~\cite{spin} 
exceeding $1/2$. At small $r_s$, however, such states are rare, and in the 
following we disregard this possibility. This allows us to neglect the 
exchange term in the interaction Hamiltonian~(\ref{2.5}). 

Keeping in mind that Eq.~(\ref{2.5}) is invariant with respect to the 
rotations of the basis of single-particle states, we can pick up the basis
in which the first term in the Hamiltonian~(\ref{2.1}) is diagonal. 
Eq.~(\ref{2.5}) then reduces to the Hamiltonian of the Constant 
Interaction Model,
\begin{equation}
\hat{H}_{dot}= 
\sum_{ns}\epsilon^\pdag_n d_{ns}^\dagger d^\pdag_{ns}
+ E_C\left({\hat N}-{\cal N}\right)^2,
\label{2.7}
\end{equation}
which is widely used in the analysis of the experimental data. We 
stress that the single-particle level spacings 
$|\epsilon_{n+1}-\epsilon_{n}|\sim\delta E$ 
here are small compared to the charging energy, and each single-particle 
state is spin-degenerate.
 
Electron transport through the dot occurs via two dot-lead junctions,
see Fig.~\ref{circuit}. As long as the conductances $G_L$ and 
$G_R$ of the left ($L$) and right ($R$) junctions are small compared to
$e^2/h$, one may use the tunneling Hamiltonian to describe the junctions:
\begin{equation}
{\hat H}_t = \sum_{\alpha k n s}t^\pdag_{\alpha kn}c^\dagger_{\alpha k s}
 d^\pdag_{ns} + {\rm H.c.},
\quad \alpha =R,L.
\label{2.8}
\end{equation}
Here $c^\dagger_{\alpha ks}$ are the creation operators of electrons 
in the leads, and $t_{\alpha kn}$ are the tunneling matrix elements 
(tunneling amplitudes) connecting the state $n$ in the dot with the state 
$k$ in the lead $\alpha$. Usually it is adequate to consider the leads 
as reservoirs of free electrons:
\begin{equation}
{\cal H}_\alpha=\sum_{ks}\xi^\pdag_{\alpha k} 
c^\dagger_{\alpha ks} c^\pdag_{\alpha ks}
\label{2.9}
\end{equation}
with continuous spectra $\xi_{\alpha k}$. In the following we will
characterize these spectra by the density of states $\nu$, same for
both leads.

The randomness of states $n$ translates into the randomness of the 
tunneling amplitudes $t_{\alpha kn}$. Similar to the single-particle 
spectrum of the dot, the tunneling amplitudes can be studied in the 
RMT framework. Regardless the details of the statistical properties 
of $t_{\alpha kn}$, the corresponding tunneling probabilities can be 
related to the average conductances $G_\alpha$ of the junctions:
\begin{equation}
\overline {|t^2_{\alpha kn}|} 
=\frac{\hbar G_\alpha}{4\pi e^2\nu}\delta E.
\label{2.10}
\end{equation}
The tunneling amplitudes determine also the rate $\Gamma_\alpha$ for 
an electron occupying a given discrete level in the dot to escape into 
the lead $\alpha=L,R$. The escape rates are determined by the tunneling 
amplitudes. With the help of Eq.~(\ref{2.10}) the average values of the 
rates are estimated  to be $\Gamma_\alpha\sim (hG_\alpha/e^2)\delta E $. 

\section{Rate equations and conductance across the dot}

At high temperatures, $T\gg E_C$, charging energy is negligible compared 
to the thermal energy of electrons. Therefore the conductance $G_\infty$ 
in this regime is not affected by charging and, independently of the 
gate voltage,
\begin{equation}
\frac{1}{G_\infty}=\frac{1}{G_L}+\frac{1}{G_R}.
\label{eq.:3.1a}
\end{equation}
At lower temperatures,
\begin{equation}
\delta E\ll T\ll E_C,
\label{3.1A}
\end{equation}
electron transport across the dot starts to depend on ${\cal N}$. The
conductance is not suppressed only within narrow regions -- Coulomb
blockade peaks, {\sl i.e.,} at the gate voltages tuned sufficiently
close to the points of charge degeneracy, 
\begin{equation}
|{\cal N} - {\cal N}^*|\lesssim T/E_C;
\label{3.1B}
\end{equation}
here ${\cal N}^*$ is a half-integer number. 

We will demonstrate this now using the method of rate equations. In 
addition to the constraint Eq.~(\ref{3.1A}), we will assume that the 
inelastic electron relaxation rate $1/\tau_\varepsilon$ within the dot 
is large compared to the escape rates $\Gamma_\alpha$. In other words, 
electron transitions between discrete levels in the dot occur before 
the electron escapes to the leads\footnote{Note that a finite inelastic 
relaxation rate requires inclusion of mechanisms beyond the Constant 
Interaction Model, {\sl e.g.,} electron-phonon collisions.}. Under 
this assumption the tunnelings across the two junctions can be 
treated independently of each other. The condition (\ref{3.1B}), 
on the other hand, allows us to take into account only two charge 
states of the dot which are almost degenerate in the vicinity of a 
Coulomb blockade peak. For ${\cal N}$ close to ${\cal N}^*$ these 
are the states with $N = {\cal N}^*\pm 1/2$ electrons on the dot. 
Hereafter we denote these states as $|N\rangle$ and $|N+1\rangle$. 
Finally, the condition (\ref{3.1A}) enables us to replace the 
discrete single-particle levels within the dot by a continuum with 
the density of states $1/\delta E$.

Applying the Fermi Golden Rule and using the described simplifications
we may write the current $I_\alpha$ from the lead $\alpha$ into the dot 
as
\begin{eqnarray}
I_{\alpha} &=&\frac{2\pi}{\hbar}\sum_{kns} |t_{\alpha k n}|^2 
\delta(\xi_k + eV_\alpha + E_N - \epsilon_n -E_{N+1}) 
\nonumber\\ 
&& ~~~~~~~~~~~~~~
\times\left\{P_{N} f(\xi_k)[1-f(\epsilon_n)] 
- P_{N+1} f(\epsilon_n)[1-f(\xi_k)]\right\}.
\nonumber
\end{eqnarray}
Here $f(\omega) = [\exp(\omega/T)+1]^{-1}$ is the Fermi function, 
$V_\alpha$ is the potential on the lead $\alpha$, see Fig.~\ref{circuit}, 
$E_{N}$ and $E_{N+1}$  are the electrostatic energies of the charge 
states $|N\rangle$ and $|N+1\rangle$, and $P_N$ and $P_{N+1}$ 
are the probabilities to find the dot in these states. Replacing the 
summations over $n$ and $k$ by integrations over the corresponding 
continua, we find
\begin{equation}
I_\alpha = \frac{G_L}{e}\left[P_N F(E_1-E_0-\mu_\alpha)
- P_{N+1} F(E_0-E_1+\mu_\alpha)\right]
\label{3.2}
\end{equation}
with $F(\omega)=\omega/[\exp(\omega/T)-1]$. In equilibrium, the current 
Eq.~(\ref{3.2}) is zero by the detailed balance. When a finite current 
flows through the junction the probabilities $P_N$ and $P_{N+1}$ deviate 
from their equilibrium values. In the steady state, the currents across 
the two junctions satisfy
\begin{equation}
I = I_L = -I_R .
\label{3.3}
\end{equation}
From Eqs.~(\ref{3.2}) and (\ref{3.3}), supplemented by the obvious 
normalization condition $P_N + P_{N+1} =1$, one finds the current across 
the dot $I$ in response to the applied bias $V = V_L - V_R$. The linear 
conductance across the dot $G=\lim_{V\to 0}{dI}/{dV}$ is then given by
\begin{equation}
G 
= G_\infty
\frac{E_C({\cal N}-{\cal N}^*)/T}{\sinh [2E_C({\cal N}-{\cal N}^*)/T]} \,.
\label{3.5}
\end{equation}
Here a half-integer ${\cal N}={\cal N}^*$ corresponds to the Coulomb 
blockade peak. In the Coulomb blockade valleys (${\cal N}\neq {\cal N}^*$), 
conductance falls off exponentially with the decreasing temperature, 
and all the valleys behave exactly the same way.

According to Eq.~(\ref{3.5}), all the Coulomb blockade peaks have 
the same height $G({\cal N}^*)=G_\infty/2$ and are symmetric about 
${\cal N} = {\cal N}^*$. Such equivalence of all the peaks occurs 
because for $T\gg \delta E$, see Eq.~(\ref{3.1A}), many energy levels in 
the dot make a significant contribution to the conductance. Decrease of 
temperature below $\delta E$ brings about a modification of Eq.~(\ref{3.5}) 
in the vicinity of the Coulomb blockade peaks 
$\left|{\cal N}-{\cal N}^*\right|\ll \delta E/E_C$. In this regime the main 
contribution to the conductance comes from tunneling through a single 
energy level in the dot. The dependence $G({\cal N}, T)$ can again be 
found from the proper rate equations with the result~\cite{ABG} 
\begin{equation}
G\propto -G_\infty \frac{\delta E}{T}\frac{df(X)/dX}{1+f(X)},
\quad\quad 
X=\frac{2E_C}{T}({\cal N}-{\cal N}^*).
\label{3.7}
\end{equation}
In writing Eq.~(\ref{3.7}) we neglected mesoscopic fluctuations of the 
peaks heights~\cite{AlhassidRMP}. Note that according to Eq.~(\ref{3.7}) 
the shape of the peak is asymmetric. The asymmetry comes about from the 
difference in the spin states of the dots in two adjacent Coulomb blockade 
valleys: Eq.~(\ref{3.7}) is written for a peak separating the valley with 
$N={\cal N}^*-1/2 =$ odd number of electrons at $X<0$ from the valley with 
$N={\cal N}^*+1/2 =$ even at $X>0$. 
The validity of Eq.~(\ref{3.7}) is restricted to the vicinity of the 
Coulomb blockade peak, $\left|{\cal N}-{\cal N}^*\right|\ll \delta E/E_C$, 
and for temperatures in the interval $\Gamma_\alpha\ll T\ll \delta E$. 
When the temperature falls below the escape rates $\Gamma_\alpha $ the 
Coulomb blockade peaks are no longer well-defined due to the onset of 
the Kondo effect.

\section{Activationless transport through a blockaded quantum dot}

According to the rate equations theory~\cite{rate}, at low
temperatures, $T\ll E_C$, conduction through the dot in the Coulomb
blockade valleys is exponentially suppressed. This suppression occurs
because the process of electron transport through the dot involves a
\textit{real transition} to the state in which the charge of the dot 
differs by $e$ from the thermodynamically most probable value. The 
probability of such fluctuation is proportional to 
$\exp[-E_C|{\cal N} -{\cal N}^*|/T]$, which explains the conductance 
suppression, see Eq.~(\ref{3.5}). Going beyond the lowest-order 
perturbation theory in conductances $G_\alpha$ allows one to consider 
processes in which states of the dot with a ``wrong'' charge 
participate in the tunneling process as \textit{virtual states}. The 
existence of these higher-order contributions to the tunneling 
conductance was first envisioned by Giaever and Zeller~\cite{Giaever}. 
The first quantitative theory of this effect, however, was developed 
much later~\cite{AN}. 

The leading contributions 
to the activationless transport, according to Refs.~\cite{AN}, are 
provided by the processes of inelastic and elastic \textit{co-tunneling}. 
Unlike the sequential tunneling, in the co-tunneling mechanism, the events 
of electron tunneling from one of the leads into the dot, and tunneling 
from the dot to the other lead occur as a single quantum process. 

\subsection{Inelastic co-tunneling}

\begin{figure}[h]
\centerline{\includegraphics[width=0.9\textwidth]{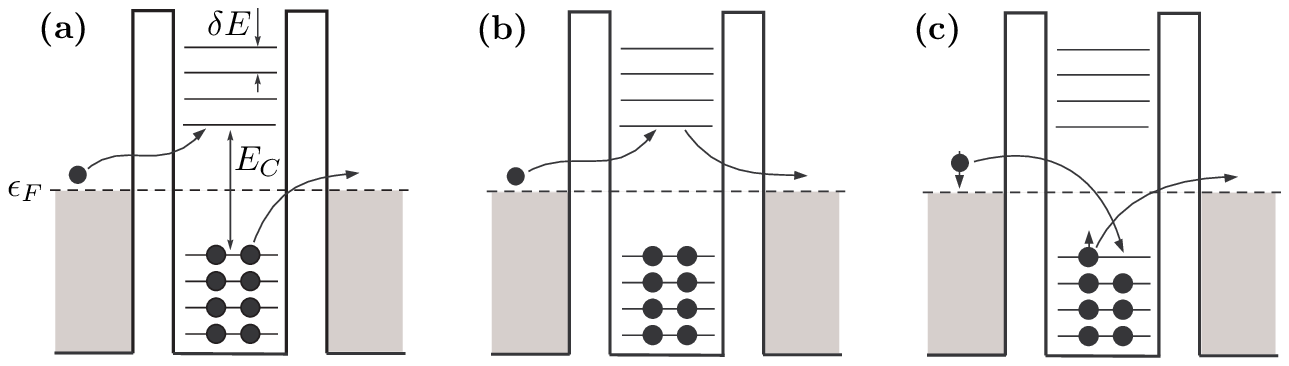}}
\vspace{2mm}
\caption{Examples of the co-tunneling processes. 
\newline (a) inelastic co-tunneling: transferring of an electron 
between the leads leaves behind an electron-hole pair in the dot;
(b) elastic co-tunneling;
(c) elastic co-tunneling with a flip of spin.
}
\label{cotunneling}
\end{figure}

In the inelastic co-tunneling mechanism, an electron tunnels from the 
lead into one of the vacant single-particle levels in the dot, while it 
is an electron occupying some other level that tunnels out of the dot, 
see Fig.~\ref{cotunneling}(a). As a result, transfer of charge $e$ between 
the leads is accompanied by a simultaneous creation of an electron-hole 
pair in the dot.

Here we will estimate the contribution of the inelastic co-tunneling to 
the conductance deep in the Coulomb blockade valley, {\sl i.e.}, at 
almost integer ${\cal N}$. Consider an electron that tunnels into the 
dot from the lead $\alpha$. If the energy of the electron relative to the 
Fermi level $\omega$ is small compared to the charging energy, 
$\omega\ll E_C$, then the energy of the virtual state involved in the
co-tunneling process is close to $E_C$. The amplitude $A_{in}$ of
the inelastic transition via this virtual state is then given by
\begin{equation}
{\cal A}^{in}_{\alpha\to \alpha'} 
= \frac{t_{\alpha kn}^*t_{\alpha' k'n'}^\pdag}{E_C}.
\label{4.1.1}
\end{equation}
The initial state of this transition has an extra electron in the
single-particle state $k$ in the lead $\alpha$, while the final state has 
an extra electron in the state $k'$ in the lead $\alpha'$ and an 
electron-hole pair in the dot (state $n$ is occupied, state $n'$ is empty).  

Given the energy of the initial state $\omega$, the number of available 
final states can be estimated from the phase space argument, familiar 
from the calculation of the quasiparticle lifetime in the Fermi liquid 
theory~\cite{Abrikosov}. For $\omega\gg\delta E$ this number is of 
the order of $(\omega/\delta E)^2$. Since the typical value of $\omega$ 
is $T$, the inelastic co-tunneling contribution to the conductance can 
be estimated as 
\[
G_{in} \sim \frac{e^2}{h} \left(\frac{T}{\delta E}\right)^2 
 \nu^2 |{\cal A}^{in}_{L\to R}|^2 .
\]
Using Eqs.~(\ref{4.1.1}) and (\ref{2.10}) we find~\cite{AN}
\begin{equation}
G_{in} \sim \frac{h}{e^2}G_LG_R\left(\frac{T}{E_C}\right)^2.
\label{4.1.2}
\end{equation}

A comparison of Eq.~(\ref{4.1.2}) with the result of the rate equations
theory Eq.~(\ref{3.5}) shows that the inelastic co-tunneling takes over
the thermally-activated hopping at moderately low temperatures
\begin{equation}
T\lesssim T_{in} =
E_C\left[\ln\left(\frac{e^2/h}{G_L+G_R}\right)\right]^{-1}.
\label{4.1.3}
\end{equation}
The smallest energy of the electron-hole pair is of the order of 
$\delta E$. At temperatures below this threshold the inelastic 
co-tunneling contribution to the conductance becomes exponentially 
small. It turns out, however, that even at much higher temperatures 
this mechanism becomes less effective than the elastic co-tunneling.

\subsection{Elastic co-tunneling}

In the process of elastic co-tunneling, transfer of charge between 
the leads is not accompanied by the creation of an electron-hole pair 
in the dot. In other words, occupation numbers of single-particle 
energy levels in the dot in the initial and final states of the co-tunneling 
process are exactly the same, see Fig.~\ref{cotunneling}(b). Here we 
will estimate the elastic co-tunneling contribution to the conductance 
near the edge of a Coulomb-blockade valley, 
\begin{equation}
{\delta E}/{E_C}\ll {\cal N}-{\cal N}^*\ll 1/2.
\label{4.2.1}
\end{equation}
Under this condition, the average number of electrons on the dot 
$N = {\cal N}^* +1/2$. A cost in electrostatic energy for a removal 
of a single electron from the dot, $E_- = 2E_C({\cal N}-{\cal N}^*)$,
see Eq.~(\ref{2.7}), is small compared to the cost to add an electron  
$E_+ \approx 2E_C$ [here we took into account the second inequality 
in Eq.~(\ref{4.2.1})]. Therefore, only hole-like virtual states with $N-1$ 
electrons on the dot contribute to the co-tunneling amplitude:
\begin{equation}
{\cal A}^{el}_{\alpha\to\alpha'} = \sum_{\epsilon_n\leq 0} 
A^n_{\alpha\to\alpha'},
\quad
A^n_{\alpha\to\alpha'} 
=  \frac{t^*_{\alpha kn}t_{\alpha'k'n}}{E_- + |\epsilon_n|} \,.
\label{4.2.2}
\end{equation}
Here $A^n$ represent amplitudes of the processes in which a hole is 
virtually created on $n$th single-particle level. Creation of a hole is only 
possible if the level is occupied, hence the restriction of the sum in 
Eq.~(\ref{4.2.2}) to the levels below the Fermi level $(\epsilon_n \leq 0)$. 

The elastic co-tunneling contribution to the conductance is 
\begin{equation}
G_{el} = \frac{e^2}{\pi\hbar} 
~
\nu^2{\left| {\cal A}^{el}_{L\to R} \right|^2}.
\label{4.2.3}
\end{equation}
If the dot-leads junctions are point contacts, then $G_{el}$
exhibits strong mesoscopic fluctuations. Indeed, tunneling matrix
elements entering Eq.~(\ref{4.2.2}) depend on the values of the
electron wave functions at the points ${\bf r}_\alpha$ of the
contacts, $t_{\alpha kn} \propto \varphi_n({\bf r}_\alpha)$. In the
RMT model, briefly discussed in Section~\ref{CIM}, the electron
wave functions in the dot $\varphi_n({\bf r})$ are random and
uncorrelated, $\overline{\varphi_n({\bf r}_\alpha) \varphi_{n'}({\bf
    r}_{\alpha'})} \propto \delta_{nn'}\delta_{\alpha\alpha'}$.
Therefore, the partial amplitudes $A^n_{L\to R}$ are random and
uncorrelated as well:
\[
\overline{\left(A^n_{L\to R}\right)^* A^m_{L\to R}} 
= \delta_{nm} \overline{\left|A^n_{L\to R}\right|^2} .
\]
Eqs.~(\ref{4.2.2}),(\ref{4.2.3}) and (\ref{2.10}) then yield for the 
average conductance 
\[
\overline{G_{el}}=\frac{e^2}{\pi\hbar}
\nu^2\sum_{\epsilon_n\leq 0} 
\overline{{\left| {\cal A}^{n}_{L\to R} \right|^2}}
\sim \frac{h}{e^2} G_L G_R \sum_{\epsilon_n\leq 0} 
\left[\frac{\delta E}{E_- + |\epsilon_n|}\right]^2.
\]
Since under the conditions (\ref{4.2.1}) the number of terms making
significant contribution to the sum over $n$ here is large, and since the 
sum is converging, one can replace the summation by an integral 
which results in~\cite{AN}
\begin{equation}
\overline{G_{el}}
\sim \frac{h}{e^2} G_L G_R  \frac{\delta E}{E_C}
\left[\frac{1}{{\cal N}-{\cal N}^*}+\frac{1}{{\cal N}^*-{\cal N}+1}\right].
\label{4.2.4}
\end{equation}
Here we have included both the hole-like and electron-like contributions 
to the conductance and restored the explicit dependence on the gate voltage 
${\cal N}$;  Eq.~(\ref{4.2.4}) is valid for 
$\left|{\cal N}-{\cal N}^* - {1}/{2}\right|\ll {1}/{2} - {\delta E}/{E_C}$.
Comparison of Eq.~(\ref{4.2.4}) with Eq.~(\ref{4.1.2}) shows that
the elastic co-tunneling dominates the electron transport already at 
temperatures
\begin{equation}
T\lesssim T_{el} = \sqrt{E_C\delta E},
\label{4.2.5}
\end{equation}
which may exceed significantly the level spacing.

Note that mesoscopic fluctuations of the elastic co-tunneling 
contribution to the conductance $G_{el}$ are of the order of its 
average and get stronger when the gate voltage is tuned closer to the 
middle of the Coulomb blockade valley~\cite{AG}. Thus, although 
$\overline{G_{el}}$ is always positive, see Eq.~(\ref{4.2.4}), the 
sample-specific value of $G_{el}$ for a given gate voltage may vanish.  

\section{Kondo regime in transport through a blockaded quantum dot}

Among the $E_C|{\cal N}-{\cal N}^*|/\delta E$ virtual states
participating in the elastic co-tunneling through a blockaded dot, 
the top-most occupied single-particle level is special. If the number 
of electrons on the dot $N$ is odd, this level is filled by a single 
electron and is spin-degenerate. Therefore the ground state of the 
dot is characterized not only by the occupations of the single-particle 
energy levels, but also by the dot's spin. This opens a possibility of
a co-tunneling process in which a transfer of an electron between the 
leads is accompanied by a flip of electron's spin with simultaneous 
flip of the spin on the dot, see Fig.~\ref{cotunneling}(c). The amplitude 
of such a process, calculated in the fourth order in tunneling matrix 
elements, diverges logarithmically when the energy $\omega$ of an 
incoming electron approaches $0$. Since $\omega\sim T$, the 
logarithmic singularity in the transmission amplitude translates into 
a dramatic enhancement of the conductance $G$ across the dot 
at low temperatures: $G$ may reach values as high as the quantum 
limit $2e^2/h$~\cite{AM, v.d.wiel}. This conductance enhancement 
is not really a surprise. Indeed, in the spin-flip co-tunneling process 
a quantum dot with odd $N$ behaves as $S=1/2$ magnetic impurity 
embedded into a tunneling barrier separating two massive conductors. 
It is known~\cite{old_reviews} since mid-60's that the presence of 
such impurities leads to zero-bias anomalies in tunneling 
conductance~\cite{classics}, which are adequately explained~\cite{AA} 
in the context of the Kondo effect~\cite{Kondo}. 

For simplicity, we will assume here that the gate voltage $\cal N$ is 
tuned to the middle of the Coulomb blockade valley with $N=\rm odd$ 
electrons on the dot. The tunneling (\ref{2.8}) mixes this state with states 
having $N\pm 1$ electrons. The electrostatic energies of these states
are high $(\sim E_C)$, hence the transitions $N\to N\pm 1$ are virtual, and 
can be taken into account perturbatively in the second order in tunneling 
amplitudes. The resulting effective Hamiltonian, valid at energies well 
below the threshold $(\sim \delta E)$ for spin excitations within the dot, 
has the form
\begin{equation} 
H_{eff} = \sum_{\alpha ks}\xi^\pdag_{k} 
c^\dagger_{\alpha ks} c^\pdag_{\alpha ks}
+ \frac{4}{E_C} \sum_{\alpha \alpha'} 
\sum_{nn'} t_{\alpha n}^* t_{\alpha' n'} 
({\bf s}_{\alpha' \alpha} \cdot {\bf S}_{nn'}).  
\label{5.1}
\end{equation}
Here
\[
{\bf s}_{\alpha \alpha'} = \sum_{kk'ss'}
c_{\alpha ks}^{\dagger}\frac{{\bm \sigma}_{ss'}}{2} c_{\alpha' k's'}^\pdag,
\quad
{\bf S}_{nn'} = {\cal P}\left[
\sum_{ss'} 
d_{ns}^{\dagger }\frac{{\bm \sigma}_{ss'}}{2}d_{n's'} ^\pdag
\right]{\cal P} ,
\]
and ${\cal P}$ is a projector onto the spin-degenerate ground state of 
an isolated dot. In the derivation of Eq.~(\ref{5.1}) we have replaced 
the tunneling amplitudes $t_{\alpha k n}$ by their values $t_{\alpha n}$ 
at the Fermi level.  In addition, we have dropped the potential scattering 
terms associated with usual elastic co-tunneling. The latter approximation 
is well justified when the conductances of the dot-leads junctions are 
small, $G_{\alpha}\ll e^2/h$, in which case $G_{el}$ is also very small, 
see Eq.~(\ref{4.2.4}). 

By $SU(2)$ symmetry, the operators ${\bf S}_{nn'}$ for any $n$ and $n'$ 
must be proportional to ${\bf S} = {\cal P}\hat{\bf S}{\cal P}$ with 
$\hat{\bf S}$ being the operator of the total spin on the dot introduced 
in Eq.~(\ref{2.6}) above: ${\bf S}_{nn'} = \lambda_{nn'} {\bf S}$.
Substitution into Eq.~(\ref{5.1}) then yields~\cite{real}
\begin{equation}
H_{eff} = \sum_{\alpha ks}\xi^\pdag_{k} 
c^\dagger_{\alpha ks} c^\pdag_{\alpha ks}
+ \sum_{\alpha \alpha'} J_{\alpha\alpha'}
({\bf s}_{\alpha'\alpha} \cdot {\bf S})
\label{5.2} 
\end{equation}
with
\begin{equation}
J_{\alpha\alpha'} = \frac{4}{E_C}
\sum_{nn'}t_{\alpha n}^* \lambda_{nn'} t_{\alpha' n'} ^\pdag.
\label{5.3} 
\end{equation}

Under the conditions of applicability of the Constant Interaction Model 
Eq.~(\ref{2.7}), the effective exchange Hamiltonian (\ref{5.2}) can be 
simplified even further. Indeed, in this case the ground state of an isolated 
dot is a singlet or a doublet, depending only on the parity of $N$. 
If $N$ is odd, then ${\bf S}$ is spin-1/2 operator. The entire spin of the 
dot is now due to the only singly occupied single-particle energy level in it 
(denoted by $n=0$ hereafter). The matrix $\lambda_{nn'}$ then reduces 
to $\lambda_{nn'} = \delta_{nn'}\delta_{n 0}$, and Eq.~(\ref{5.3}) yields  
\begin{equation}
J_{\alpha\alpha'} = \frac{4}{E_C} t_{\alpha 0}^* t_{\alpha' 0} \,.
\label{5.4} 
\end{equation}
The $2\times 2$ Hermitian matrix $\hat J$ with elements given by 
Eq.~(\ref{5.4}) has an important property: since $\det \hat J =0$, 
one of its eigenvalues vanishes, while the remaining eigenvalue,
\begin{equation}
J = {\rm Tr} \hat J = \frac{4}{E_C}\left(|t_{L0}^2| + |t_{R0}^2|\right),
\label{5.4A} 
\end{equation}
is strictly positive. By an appropriate rotation in the $R-L$ 
space the Hamiltonian (\ref{5.2}) can then be brought into the 
"block-diagonal`` form 
\begin{equation}
H_{eff} = H[\varphi] + H[\psi] 
\label{5.5}
\end{equation}
with 
\begin{eqnarray}
H[\varphi] &=& \sum_{ks}\xi^\pdag_{k} \varphi^\dagger_{ks} \varphi^\pdag_{ks},
\label{5.6.A} \\
H[\psi] &=& \sum_{ks}\xi^\pdag_{k} \psi^\dagger_{ks} \psi^\pdag_{ks}
+ J({\bf s}_{\psi} \cdot {\bf S}) .
\label{5.6.B} 
\end{eqnarray}
Here ${\bf s}_{\psi} = \sum_{kk'ss'}\psi_{ks}^{\dagger } \left({{\bm
      \sigma }_{ss'}}/{2}\right) \psi_{k's'}^\pdag$, and the operators
$\psi$ and $\varphi$ are certain linear combinations of the original
operators $c_{R,L}$. 

To get an idea about the physics of the Kondo model, let us first replace 
the operator ${\bf s}_\psi$ in Eq.~(\ref{5.6.B}) by spin-1/2 operator 
${\bf S}_\psi$. The ground state of the resulting Hamiltonian of two spins
$\widetilde{H} = J({\bf S}_\psi \cdot {\bf S})$ is obviously a singlet. 
The excited state (a triplet) is separated from the ground state by the 
energy gap $J>0$. This separation can be interpreted as the binding energy 
of the singlet. Unlike ${\bf S}_\psi$ in this simple example, the operator 
${\bf s}_\psi$ in Eq.~(\ref{5.6.B}) is merely a spin density of the 
conduction electrons at the site of the "magnetic impurity``. Because 
conduction electrons are freely moving in space, it is hard for the impurity 
to "capture`` an electron and form a singlet. Yet, even a weak local exchange 
interaction suffices to form a singlet ground state~\cite{PWA_book,Wilson}.
However, the binding energy of this singlet is given not by the exchange 
constant $J$, but by the so-called Kondo temperature
\begin{equation}
T_K \sim \delta E \exp({-1/\nu J}).
\label{5.6A}
\end{equation}
With the help of Eqs.~(\ref{5.4A}) and (\ref{2.10}) one obtains from 
Eq.~(\ref{5.6A}) the estimate
\begin{equation}
\ln\left(\frac{\delta E}{T_K}\right)
\sim \frac{1}{\nu J}
\sim \frac{e^2/h}{G_L + G_R}\frac{E_C}{\delta E}.
\label{5.6}
\end{equation}
Since $G_\alpha\ll e^2/h$, and $E_C\gg \delta E$, the r.h.s. of 
Eq.~(\ref{5.6}) is a product of two large parameters. Therefore, the 
Kondo temperature $T_K$ is small compared to the mean level spacing in 
the dot:
\begin{equation}
T_K \ll \delta E.
\label{5.7}
\end{equation}
It is this separation of the energy scales that justifies the use of the 
effective low-energy Hamiltonian (\ref{5.1}) for the description of 
the Kondo effect in a quantum dot system. The inequality (\ref{5.7}) 
remains valid even if the conductances of the dot-leads junctions 
$G_\alpha$ are of the order of $2e^2/h$. However, in this case 
neither Eq.~(\ref{5.4A}) nor the estimate Eq.~(\ref{5.6}) are 
applicable~\cite{GHL}. 

As it follows from Eqs.~(\ref{5.6.A}) and (\ref{5.6.B}), one of the
"channels`` $(\varphi)$ of conduction electrons completely
decouples\footnote{It should be emphasized that the decoupling is
characteristic of the Constant Interaction Model rather than a
generic property. In general, both electronic channels are coupled
to the dot.  This, however, does not lead to qualitative changes in
the results if spin of the dot is $1/2$, see~\cite{real}.} from the
dot, while the $\psi$-particles are described by the standard
single-channel antiferromagnetic Kondo model~\cite{Kondo}. 
Therefore, the thermodynamic properties of a quantum dot in the 
Kondo regime are identical to those of the conventional Kondo problem 
for a single magnetic impurity in a bulk metal; thermodynamics of the 
latter model is fully studied~\cite{bethe}.  However, all the experiments
addressing the Kondo effect in quantum dots test their transport
properties rather then thermodynamics. The electron current operator is not
diagonal in the $(\psi,\phi)$ representation, and the contributions of
these two sub-systems to the conductance are not additive. In the
following we establish the relations of linear conductance and, in some 
special case, of the non-linear differential conductance as well, 
to the t-matrix of the conventional Kondo problem.

\subsection{Linear response}

The linear conductance can be calculated from the Kubo formula
\begin{equation}
G = \lim_{\omega\to 0} \frac{e^2}{\hbar}\frac{1}{\omega}
\int_0^\infty dt e^{i\omega t} \left\langle \left[j(t), j(0) \right]\right\rangle ,
\label{5.1.1} 
\end{equation}
where the particle current operator $j$ is given by
\begin{equation}
j = \frac{d}{dt} \frac{1}{2} \left(\hat N_R - \hat N_L\right).
\label{5.1.2} 
\end{equation}
Here $\hat N_\alpha = \sum_{ks}^\pdag c^\dagger_{\alpha ks}c^\pdag_{\alpha ks}$ 
is the operator of the total number of electrons in the lead $\alpha$.
In order to take the full advantage of the decomposition (\ref{5.5})-(\ref{5.6.B}), 
we need to rewrite $j$ in terms of the operators $\psi,\varphi$. These
operators are related to the original operators $c_{R,L}$ representing the 
electrons in the right and left leads via
\begin{equation}
\left(\begin{array}{c}
\psi_{ks} \\ 
\varphi_{ks}
\end{array}\right)
= \hat U^*
\left(\begin{array}{c}
c_{Rks} \\ 
c_{Lks}
\end{array}\right),
\label{5.1.3} 
\end{equation}
where $\hat U$ is $2\times 2$ unitary matrix that diagonalizes the matrix 
of the exchange constants (\ref{5.3}): 
$\hat U\hat J \hat U^\dagger = {\rm diag}\{J,0\}$. The matrix $\hat U$ 
can be parametrized as 
\begin{equation}
\hat U = e^{i\theta_0 \tau_y}e^{i \phi_0 \tau_z},
\label{5.1.4} 
\end{equation}
where $\tau_i$ are the Pauli matrices acting in the R-L space 
($\tau_+ = \tau_x + i\tau_y$ transforms $L$ to $R$) and the angle 
$\theta_0$ satisfies $\tan \theta_0 = |t_{L0}/t_{R0}|$.
It follows from Eqs.~(\ref{5.1.3}) and (\ref{5.1.4}) that, independently of $\phi_0$,
\begin{equation}
\hat N_R - \hat N_L = \cos(2\theta_0) \left(\hat N_\psi - \hat N_\varphi\right) 
- \sin(2\theta_0) \sum_{ks}
\left(
\psi_{ks}^\dagger \varphi_{ks}^\pdag + \varphi_{ks}^\dagger \psi_{ks}^\pdag
\right) .
\label{5.1.5} 
\end{equation}
Note that the current operator $j$ entering the Kubo formula (\ref{5.1.1}) is to 
be calculated with the equilibrium Hamiltonian (\ref{5.5})-(\ref{5.6.B}). Since 
both $\hat N_\psi$ and $\hat N_\varphi$ commute with $H_{eff}$, the first term in 
Eq.~(\ref{5.1.5}) makes no contribution to $j$. When the second term in 
(\ref{5.1.5}) is substituted into Eq.~(\ref{5.1.2}) and then into Eq.~(\ref{5.1.1}), 
the result, after integration by parts, can be expressed via 2-particle correlation 
functions such as 
$\left\langle \psi^\dagger(t)\varphi(t) \varphi^\dagger(0)\psi(0)\right\rangle$. 
Due to the block-diagonal structure of $H_{eff}$, see Eq.~(\ref{5.5}), these 
correlation functions factorize into products of single-particle correlation functions 
describing the (free) $\varphi$-particles and the (interacting) $\psi$-particles. 
The result of the evaluation of the Kubo formula can then be written 
as\footnote{Further details about this calculation can be found, {\sl e.g.}, in the 
Appendix B of Ref.~\cite{PG}.}
\begin{equation}
G = G_0 \int d\omega \left(-{df}/{d\omega}\right) 
\frac{1}{2} \sum_s \left[- \pi\nu {\rm Im}~ T_s (\omega)\right]. 
\label{5.1.6}
\end{equation}
Here 
\begin{equation}
G_0 = \frac{2e^2}{h}  \sin^2(2\theta_0) 
= \frac{2e^2}{h} \left|\frac{2t_{L0} t_{R0}}{|t_{L0}^2|+|t_{R0}^2|}\right|^2,
\label{5.1.6A}
\end{equation}
\noindent 
$f(\omega)$ is the Fermi function, and $T_s$ is the t-matrix for the Kondo 
model Eq.~(\ref{5.6.B}). The t-matrix is related to the exact retarded 
Green function of the $\psi$-particles in the usual way:
\begin{equation}
G_{ks,k's'}(\omega) = G^0_{k}(\omega) 
+ G^0_{k}(\omega) \left[\delta_{ss'} T_s(\omega)\right] G^0_{k'}(\omega), 
\label{5.1.7}
\end{equation}
Here $G^0_{k}(\omega) = (\omega - \xi_k +i0)^{-1}$  and $G_{ks,k's'}(\omega)$ 
is the Fourier transform of 
\[
G_{ks,k's'}(t) = -i\theta(t)
\left\langle\left\{\psi_{ks}^\pdag(t),\psi_{k's'}^\dagger\right\}\right\rangle,
\]  
where $\langle ...\rangle$ stands for the thermodynamic averaging with 
the Hamiltonian (\ref{5.6.B}). In writing Eq.~(\ref{5.1.7}) we took into 
account the conservation of the total spin (which implies that $G_{ks,k's'}$ 
is diagonal in $s,s'$) and the fact that the interaction in Eq.~(\ref{5.6.B}) 
is local (which in turn means that the t-matrix is independent of $k$ and $k'$). 

\subsubsection{Weak coupling regime: $T_K\ll T\ll \delta E$}

Kondo effect becomes important at low temperatures $T\sim T_K \ll \delta E$. 
The exchange term in the Hamiltonian~(\ref{5.6.B}), however, describes 
transitions between the electronic states within the band of the width 
$2D_0$ with $D_0\sim \delta E$ centered at the Fermi level $\epsilon_F$. 
The transitions between the states close to $\epsilon_F$ and the states within 
a narrow strip of energies of the width $\delta D$ near the edges of the band
 are associated with high energy deficit $\sim \delta E \gg T$. Hence, these 
transitions are virtual and their influence on the states near $\epsilon_F$ can 
be taken into account perturbatively in the second order. This yields an
effective Hamiltonian acting within the band of a reduced width 
$D = D_0-\delta D$, which turns out to have the same form as 
Eq.~(\ref{5.6.B}), but with a modified value of the exchange amplitude 
$J$~\cite{PWA}.  Successive reductions of the bandwidth by small steps 
$\delta D$ can be viewed as a continuous process during which the initial 
Hamiltonian Eq.~(\ref{5.6.B}) is transformed to an effective low-energy 
Hamiltonian acting within the band of the width $D\ll D_0$. The evolution 
of the exchange amplitude during this transformation (known as the poor 
man's scaling~\cite{PWA}) can be cast into the form of an equation
\[
\frac{dJ}{d\zeta} =\nu J^2,
\quad
\zeta = \ln\left({D_0}/{D}\right).
\]
The renormalization described by this equation can be continued until 
the bandwidth $D$ becomes of the order of the temperature $T$ or 
relevant electron energy $|\omega|$. The resulting effective exchange 
constant is $\omega$- and $T$-dependent~\cite{PWA,AAA},
\begin{equation}
\nu J(\omega,T) = \left[\ln\frac{\max\{|\omega|,T\}}{T_K}\right]^{-1}.
\label{5.1.8}
\end{equation}
Calculation of the t-matrix in the lowest (second) order in $\nu J(\omega,T)$ 
results in
\begin{equation}
-\pi\nu {\rm Im} T_s(\omega) = \frac{3\pi^2}{16} \left[\nu J(\omega,T)\right]^2 .
\label{5.1.9}
\end{equation}
Substitution of (\ref{5.1.9}) and (\ref{5.1.8}) into Eq.~(\ref{5.1.6}) and 
evaluation of the integral over $\omega$ with the logarithmic accuracy then 
yield for the conductance across the dot
\begin{equation}
G = G_0 \frac{3\pi^2/16}{\left[\ln(T/T_K)\right]^2}\,, 
\quad
T_K \ll T\ll \delta E .
\label{5.1.10}
\end{equation}
Corrections to Eq.~(\ref{5.1.10}) contain higher powers of $1/\ln(T/T_K)$. 
Thus, it is often said that Eq.~(\ref{5.1.10}) represents the conductance 
in the {\it leading logarithmic approximation}.

\subsubsection{Strong coupling regime: $T\ll T_K$}

As temperature approaches $T_K$, the leading logarithmic approximation 
result diverges, see Eq.~(\ref{5.1.10}). This divergence signals the failure 
of the approximation, as the conductance in any case can not grow higher 
than $2e^2/h$. To obtain a more precise bound, we consider in this section 
the conductance in the strong coupling regime $T\ll T_K$. 

We start with the zero-temperature limit $T=0$. Since, as discussed 
above, the ground state of the Hamiltonian~(\ref{5.6.B}) is not degenerate, 
the scattering of conduction electrons by a quantum dot is completely 
characterized by the scattering phase shifts $\delta_s$ for electrons with 
spin $s$ at the Fermi level. The t-matrix is then given by the standard 
scattering theory expression
\begin{equation}
-\pi\nu T_s(0) = \frac{1}{2i}\left(e^{2i\delta_s}-1\right) 
= e^{i\delta_s} \sin\delta_s. 
\label{5.1.11}
\end{equation}

In order to find the two phase shifts $\delta_s$, we need two independent 
relations. The first one follows from the invariance of the Kondo 
Hamiltonian~(\ref{5.6.B}) under the particle-hole transformation 
$\psi^\pdag_{ks} \to s\psi^\dagger_{-k,-s}$ (here $s=\pm 1$ for 
spin-up/down electrons and $k$ is counted from $k_F$). The particle-hole 
symmetry implies the relation for the t-matrix
\[
T_s(\omega) = - T^*_{-s}(-\omega),
\]
valid at all $\omega$ and $T$. In view of Eq.~(\ref{5.1.11}), it translates 
into the relation for the phase shifts at the Fermi level ($\omega = 0$):
\begin{equation}
\delta_s + \delta_{-s} = 0~({\rm mod}~\pi) .
\label{5.1.12A}
\end{equation}
Note that, as it is obvious from Eq.~(\ref{5.1.11}), the phase shifts 
are defined only modulo $\pi$ (that is, $\delta_s$ and $\delta_s +\pi$ 
describe equivalent scattering states). This ambiguity can be removed 
by setting the values of the phase shifts corresponding to $J = 0$ in 
Eq.~(\ref{5.6.B}) to zero. With this convention, Eq.~(\ref{5.1.12A})
becomes
\begin{equation}
\delta_\uparrow + \delta_\downarrow = 0.
\label{5.1.12B}
\end{equation}

The second relation follows from the fact that the ground state 
of our system is a singlet at $J>0$. In the absence of exchange
($J=0$) and at $T=0$, an infinitesimally small magnetic field acting on
the dot's spin, 
\[
\delta H[\psi] = BS^z, \quad B\to +0,
\] 
would polarize it. Since the free electron gas has zero spin in the
ground state, the total spin in any large but finite region of space 
$\cal V$ surrounding the dot coincides with the spin of the dot, 
$\langle S^z\rangle_{J=0} = -1/2$. If the exchange with electron gas 
is now turned on, $J > 0$, the infinitesimally weak field will not prevent
the formation of a singlet ground state. In this state, the total spin
within the region $\cal V$ is zero. Such change of the spin is possible 
only if the numbers of spin-up and spin-down electrons in this region 
have changed in order to compensate for the spin of the dot:
\begin{equation}
\delta N_\uparrow -\delta N_\downarrow = 1, 
\quad
\delta N_s = \langle \hat N_s \rangle_{J>0} - \langle \hat N_s \rangle_{J=0}.
\label{5.1.13A}
\end{equation}
Here $\hat N_s $ are operators of the number of electrons with spin $s$ 
within the region $\cal V$. By the Friedel sum rule, $\delta N_s$ are related 
to the scattering phase shifts: $\delta N_s = \delta_s/\pi $. 
Eq.~(\ref{5.1.13A}) then gives
\begin{equation}
\delta_\uparrow - \delta_\downarrow = \pi.
\label{5.1.13B}
\end{equation}
Combining Eqs.~(\ref{5.1.12B}) and (\ref{5.1.13B}), we find
$|\delta_s| = \pi/2$.
Eqs.~(\ref{5.1.6}) and (\ref{5.1.11}) then yield for zero-temperature 
conductance across the dot~\cite{AM}
\begin{equation}
G(0) = G_0\frac{1}{2} \sum_s \sin^2\delta_s = G_0.
\label{5.1.14}
\end{equation}
Thus, the grows of the conductance across the dot with lowering the 
temperature is limited only by the value of $G_0$. This value, see 
Eq.~(\ref{5.1.6A}), depends only on the ratio of the tunneling amplitudes 
$|t_{L0}/t_{R0}|$ between the leads and the last occupied single-particle 
energy level in the dot. If $|t_{L0}|=|t_{R0}|$, the conductance at $T=0$
will reach the maximal value $G=2e^2/h$ allowed by quantum 
mechanics~\cite{AM}.

The ground state of the Kondo Hamiltonian~(\ref{5.6.B}) is a singlet 
formed by the spin of the dot and a spin made up of the spins of 
conduction electrons.  Finite-temperature correction to Eq.~(\ref{5.1.15}) 
can be found by considering virtual transitions from the singlet ground state 
to excited states in which the singlet is broken up~\cite{N}. The transitions 
can be studied by an expansion in inverse powers of the singlet binding 
energy $T_K$. The reader is referred to the original papers~\cite{N} for 
the details about this approach. Here we just quote the result for the 
imaginary part of the t-matrix~\cite{AL}:
\begin{equation}
-\pi\nu {\rm Im}~ T_s (\omega) = 1-\frac{3\omega^2 +\pi^2T^2}{2T_K^2}\,,
\quad
|\omega|,T\ll T_K
\label{5.1.15}
\end{equation}
Substitution of Eq.~(\ref{5.1.15}) into Eq.~(\ref{5.1.6}) yields
\begin{equation}
G = G_0\left[1 - \left({\pi T}/{T_K}\right)^2\right],
\quad
T\ll T_K
\label{5.1.16}
\end{equation}
Accordingly, corrections to the conductance are quadratic in temperature -- 
a typical result for the Fermi liquid theory~\cite{N}.

The weak-coupling ($T\gg T_K$) and the strong-coupling ($T\ll T_K$)
asymptotes of the conductance have a strikingly different structure.
Nevertheless, since the Kondo effect is a crossover phenomenon rather
than a phase transition~\cite{bethe,PWA_book,Wilson}, the dependence
$G(T)$ is a smooth and featureless~\cite{Costi} function throughout
the crossover region $T\sim T_K$.

Finally, note that both Eqs.~(\ref{5.1.10}) and (\ref{5.1.16}) have
been obtained here for the particle-hole symmetric
model~(\ref{5.6.B}). This approximation is equivalent to neglecting
the elastic co-tunneling contribution to the conductance $G_{el}$. The
asymptotes (\ref{5.1.10}),(\ref{5.1.14}), however, remain
qualitatively correct~\cite{real} as long as $G_{el}/ G_0\ll 1$. The
overall temperature dependence of the linear conductance in the Coulomb
blockade valley in the presence of the Kondo effect is sketched in Fig.~3.

\begin{figure}[h]
\centerline{\includegraphics[width=0.8\textwidth]{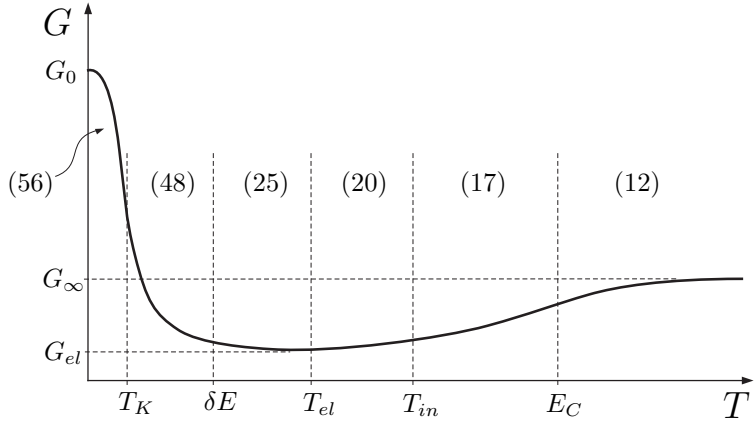}}
\vspace{2mm}
\caption{Schematic temperature dependence of the conductance in 
the middle of the Coulomb blockade valley with $N=\rm odd$ electrons 
on the dot. The numbers in brackets refer to the corresponding 
equations in the text.
}
\label{overall}
\end{figure}

\subsection{Beyond linear response}
\label{NONEQ}

In order to study transport through a quantum dot away from 
equilibrium we add to the effective Hamiltonian (\ref{5.5})-(\ref{5.6.B}) 
a term 
\[
H_V = \frac{eV}{2} \left(\hat N_L - \hat N_R\right)
\]
describing a finite voltage bias $V$ applied between the left and right 
electrodes. Here we will evaluate the current across the dot at arbitrary 
$V$ but under the simplifying assumption that the dot-lead junctions are 
strongly asymmetric: $G_L\ll G_R$. Under this condition the angle 
$\theta_0$ in Eq.~(\ref{5.1.4}) is small, 
\[
\theta_0\approx |t_{L0}/t_{R0}|\ll 1.
\] 
Expanding Eq.~(\ref{5.1.5}) to the linear order in $\theta_0$ we obtain
\begin{eqnarray}
H_V(\theta_0) &=& \frac{eV}{2} \left(\hat N_\varphi - \hat N_\psi\right) 
+ eV \theta_0 \hat A,
\label{5.2.2} \\ 
\hat A &=& \sum_{ks}
\varphi_{ks}^\dagger \psi_{ks}^\pdag + {\rm H.c.} 
\nonumber
\end{eqnarray}
The first term in the r.h.s. of Eq.~(\ref{5.2.2}) describes the voltage 
bias between the reservoirs of $\varphi$ and $\psi$-particles, while the 
second term has an appearance of the $k$-conserving ``tunneling" with  
$V$-dependent ``tunneling amplitude". These two contributions 
correspond to the two terms in the r.h.s. of Eq.~(\ref{5.1.5}). 
Similarly, the current operator $j$, see Eq.~(\ref{5.1.2}), can be also 
split into two contributions:
\begin{eqnarray}
j &=& j_0 + \delta j, 
\label{5.2.3} \\
 j_0 &=& \frac{i}{2}\left[H_{eff} + H_V, \hat N_\psi - \hat N_\varphi \right]
= i eV \theta_0 \sum_{ks}\varphi_{ks}^\dagger \psi_{ks}^\pdag 
+ {\rm H.c.} ,
\nonumber\\
\delta j &=& - \theta_0 \frac{d}{dt} \hat A .
\nonumber
\end{eqnarray}
It turns out that $\delta j$ in Eq.~(\ref{5.2.3}) makes no contribution 
to the average current in the leading (second) order in $\theta_0$. 
Indeed, in this order 
\[
\langle \delta j\rangle = 
i\theta_0^2 eV \frac{d}{dt} \int^t_{-\infty}dt' 
\left\langle [\hat A(t), \hat A(t')] \right\rangle_0 ,
\]
where $\langle ...\rangle_0$ denotes thermodynamic averaging with 
the Hamiltonian $H_0 = H_{eff} + H_V(0)$. The thermodynamic
(equilibrium) averaging is well defined here because the ``tunneling" 
term in Eq.~(\ref{5.2.2}) is absent at $\theta_0 =0$. Taking into 
account that $\left\langle [\hat A(t'), \hat A(t)] \right\rangle_0$ depends only 
on $t-t' =\tau$, we find 
\[
\langle \delta j\rangle = i\theta_0^2 eV \frac{d}{dt} \int^\infty_0d\tau 
\left\langle [\hat A(\tau), \hat A(0)] \right\rangle_0 
\equiv 0.
\]
To the contrary, the term $j_0$ in Eq.~(\ref{5.2.3}) makes a finite 
contribution to the average current. Evaluation of this contribution in 
the leading order in $\theta_0$ proceeds similarly to the calculation of 
the tunneling current between two massive conductors (see, 
{\sl e.g.},~\cite{Mahan}), and yields for the differential conductance 
across the dot
\begin{equation}
{dI}/{dV} = G_0 \int d\omega \left(-{df}/{d\omega}\right) 
\frac{1}{2} \sum_{s} \left[- \pi\nu {\rm Im}~ T_s (\omega + eV)\right] 
\label{5.2.4}
\end{equation}
with 
\[
G_0 = \frac{2e^2}{h}(2\theta_0)^2 
\approx \frac{8e^2}{h}\left|\frac{t_{L0}}{t_{R0}}\right|^2.
\] 
Note that the $V\to 0$ limit of Eq.~(\ref{5.2.4}) coincides with the 
small $\theta_0$ limit of the linear response result Eq.~(\ref{5.1.6}).
Using now Eqs.~(\ref{5.1.8}),(\ref{5.1.9}),(\ref{5.1.15}), and (\ref{5.2.4}) 
we obtain for the differential conductance at $T\ll eV\ll \delta E$
\begin{equation}
\frac{dI}{dV}= \left\{ 
\begin{array}{ll}
G_0 \left[1-\displaystyle{\frac{3}{2}\left(\frac{eV}{T_K}\right)^2}\right], 
& eV\ll T_K  
\\ \\
G_0\displaystyle{\frac{3\pi^2/16}{\left[\ln(eV/T_K)\right]^2}}~, 
& eV \gg T_K 
\end{array}
\right.
\label{5.2.5}
\end{equation}
Thus, a large voltage bias $eV\gg T$ has qualitatively the same destructive 
effect on the Kondo physics as the temperature does. If temperature $T$ 
exceeds the bias, $T\gg eV$, the differential conductance $dI/dV$ coincides 
with the linear conductance $G$, see Eqs.~(\ref{5.1.10}),(\ref{5.1.16}) above.

\section{Conclusion}

In the simplest form of the Kondo effect considered in this article,
a quantum dot behaves essentially as a magnetic impurity with spin 
$1/2$ coupled via exchange interaction to two conducting 
leads~\cite{AA}. However, the characteristic energy scale for the 
intra-dot excitations is much smaller than the corresponding scale 
for a real magnetic impurity. This allows one to induce some 
degeneracies in the ground state of a dot which are more exotic than 
just the spin degeneracy considered above. One of the possibilities is 
to create a degeneracy between a singlet state and a component of a 
triplet state by applying a magnetic field to a dot with an even number 
of electrons. Depending on the relation between the Zeeman energy 
and the value of the diamagnetic shift of orbital levels in a magnetic 
field, this degeneracy gives rise to various ``flavors'' of the Kondo 
effect. A review of the corresponding experiments and theory can be 
found in~\cite{induced_review}

In our discussion of the Kondo effect we assumed that the dot remains
close to an equilibrium state even under the conditions of an electron
transport experiment. In fact, this limitation led us to consider a
very asymmetric setup of a quantum dot device in Sec.~\ref{NONEQ},
where the non-linear electron transport is discussed. One of the
advantages of quantum dots, however, is that these devices allow one
to study the Kondo effect in truly out-of-equilibrium conditions. Such
conditions may be created by applying a significant dc bias between
the leads, or by irradiating a dot by microwaves~\cite{Elzerman}. In
the latter case one can monitor the ``health'' of the Kondo effect by
measuring the linear dc conductance at the same time. It turns out
that microwaves suppress the effect~\cite{Elzerman} by destroying the
coherence of the spin state of the dot~\cite{KNG}. In the former case,
one can apply a strong magnetic field in addition to biasing the
dot. Zeeman splitting then leads to peaks in the differential
conductance at a finite bias ~\cite{Appelbaum,MWL}. Interest to this
problem was recently revived, see, {\sl e.g.}, \cite{Rosch} and
references therein.

These are only a few out of many possible extensions of the simple 
model discussed in this review.

\section*{Acknowledgements}
This work is supported by NSF grants DMR97-31756, DMR02-37296, and EIA02-10736.

\end{document}